\documentclass{article}
\usepackage{xcolor}
\usepackage{graphicx} 
\usepackage{etaremune}
\usepackage{mathrsfs}
\usepackage{complexity}
\usepackage{bm}
\usepackage{amsmath}
\usepackage{amsthm}
\usepackage{amssymb}
\usepackage{stmaryrd}
\usepackage{mathtools}
\usepackage{changepage} 
\usepackage{enumitem}
\usepackage{url}
\usepackage{authblk}
\usepackage[many]{tcolorbox}
\usepackage{float}
\usepackage{natbib}
\usepackage{color}
\usepackage{tikz,lipsum,lmodern}
\usepackage{dcolumn}
\usepackage{tabularx}

\title{Demographic synchrony increases the vulnerability of human societies to collapse}

\author[1,2,3,*]{Marcus J. Hamilton}
\author[4,5]{Robert S. Walker}

\affil[1]{Department of Anthropology, University of Texas at San Antonio, San Antonio, 78249, TX, USA}
\affil[2]{School of Data Science, University of Texas at San Antonio, San Antonio, 78249, TX, USA}
\affil[3]{Santa Fe Institute, Santa Fe, 87501, NM, USA.}
\affil[4]{Department of Anthropology, University of Missouri, Columbia, MO, USA}
\affil[5]{Center for Applied Statistics and Data Analysis, University of Missouri, Columbia, MO, USA}

\date{\today}

\begin{document}
\maketitle

\pagebreak



\pagebreak

\begin{abstract}
Why do human populations remain vulnerable to collapse, even when they are large? Classical demographic theory predicts that volatility in growth should decline rapidly with size due to the averaging effects of the law of large numbers. As such, while small-scale societies may be demographically fragile, large-scale societies should be much more stable. Using a large census dataset of 200+ indigenous societies from Brazil, we show that this prediction does not hold. Instead of volatility declining as the square root of population size, it falls much more slowly. This means that individuals within communities do not behave as independent demographic units as their lives are correlated through cooperation, shared subsistence practices, overlapping land use, and exposure to common shocks such as disease outbreaks or failed harvests. These correlations build demographic synchrony, drastically reducing the effective demographic degrees of freedom in a population, keeping volatility higher than expected at all scales. As a result, large-scale populations fluctuate as if they were much smaller, increasing their vulnerability to collapse. This helps explain why human societies of all sizes seem vulnerable to collapse, and why the archaeological and historical record is filled with examples of large, complex societies collapsing despite their size. We suggest demographic synchrony provides a general mechanism for understanding why human populations remain vulnerable across all scales: Scale still stabilizes synchronous populations via density increases, but synchrony ensures that stability grows only slowly with size, leaving large populations more volatile — and more vulnerable — than classical demographic theory predicts.
\end{abstract}

\pagebreak

\section{Introduction}

The human past is populated by societies and civilizations that grew in scale over time only to collapse from some combination of internal and external mechanisms \cite{diamond2011collapse, tainter1988collapse, walker2011social, hamilton2014crash, walker2014amazonian, currie2010rise}. Indeed, societal collapse is a central process in cultural evolution and plays a crucial role in human demographic history \cite{hamilton2018stochastic, gurven2019periodic, butzer2012collapse, cumming2017unifying, turchin2018historical}. The drivers, causes, and consequences of societal collapse are widely debated, and individual instances have developed into their own fields of study, whether the 17th century CE collapse of Easter Island \cite{dinapoli2020model}; the collapse of Late Classic Maya city-states in the 1st millennium CE \cite{haug2003climate}; the Late Bronze Age societal collapse across the Mediterranean basin during the 12th century BCE \cite{drake2012influence}; or the fall of the Western Roman Empire in the 5th century CE \cite{heather2005fall}. 

Throughout this paper we use the term collapse in a demographic sense to mean a sudden and often irreversible reduction in population viability. Collapse occurs when demographic fluctuations drive populations below thresholds from which recovery is unlikely, often captured by the concept of a minimum viable population size \cite{dennis1991estimation, morris2002quantitative, lande2003stochastic, hamilton2018stochastic}. Our definition is narrower than the archaeological or historical use of the term, which often refers to the breakdown of social and political institutions in general \cite{tainter1988collapse, turchin2018historical}. However, there is a sense in which all societal collapse events, no matter how instigated, are ultimately demographic in nature if institutional failures result in a loss of population viability. Here, we focus on these demographic mechanisms, specifically how correlated demographic outcomes create demographic synchrony inflating volatility and increasing the probability of extinction-like outcomes even in populations that appear numerically large.

We ask why demographic scale fails to provide the buffer against vulnerability that theory predicts. Classical demographic models assume that births, deaths, and migrations are independent stochastic events \cite{lande2003stochastic, keyfitz2005applied}. Under this baseline, volatility in per-capita growth rates declines with the square root of population size, $\sigma(r)\propto N^{-1/2}$, so that larger populations should be steadily---and rapidly---more stable, as guaranteed by the law of large numbers. Yet the archaeological record shows repeated collapse of societies across all demographic scales, from hunter-gatherer populations to geographically vast empires, suggesting that size alone is not the guarantor of resilience. A fundamental theoretical challenge is to understand how intra-population correlations in demographic outcomes alter demographic scaling at the population scale and increase the vulnerability of human societies.

Here we argue that demographic synchrony---the correlation of demographic outcomes across individuals within populations---systematically inflates volatility by reducing the effective number of independent demographic draws. In other words, synchrony reduces the demographic degrees of freedom in a population by introducing correlations that violate assumptions of independence. In a synchrony-extended model, populations behave as though they are composed not of $N$ independent individuals, but of $K(N)$ effective demographic units, where $K$ grows more slowly than $N$. As synchrony increases, the law of large numbers weakens as variance is no longer averaged away, and volatility remains high even in large-scale populations. This insight provides a theoretical bridge between human demography and ecological models of synchrony, in which correlated trajectories across subpopulations elevate regional extinction risk \cite{bjornstad1999spatial, koenig2002global, liebhold2004spatial, ims1990ecology, earn1998persistence, brown1977turnover}. 

Moreover, we explain how this phenomena introduces a demographic tension in human societies that we term the cooperation--synchrony paradox: cooperation between individuals evolves in the human species to reduce risk in fitness-related outcomes (such as foraging, provisioning and childcare) and by introducing economies of scale, thus solving adaptive problems through scale-limited collective action \cite{hamilton2007nonlinear}. But the mathematical consequence of averaging over stochasticity through cooperation is to build correlations within populations thus building synchrony. Synchrony has the effect of making populations more vulnerable to stochastic shocks.

However, it has been difficult to study such dynamics empirically in traditional human populations because long-term demographic data for small-scale societies are scarce. This is because human population dynamics fall into an anthropological blind spot; too long to be captured by ethnographic studies and too short to be captured in archaeological or paleoanthropological contexts. Moreover, population dynamics themselves are not preserved in the archaeological record (though these are often inferred from material correlates), and capturing time series data of the length and replication---dozens of societies measured over multiple generations---required for statistical models are simply not logistically feasible from field-based studies. As such, human demography is most often studied through census data or historical records, which rarely include detailed data on indigenous populations. 

To address this gap, we assemble and analyze more than 200 time series of indigenous Amazonian populations spanning multiple regions and decades. These data describe the population dynamics of a total of 228 indigenous forager-horticulturalist groups in the Brazilian Amazon compiled from censuses undertaken by the Brazilian government, some with estimates beginning in the 18th century. The time series are short, sparse, and uneven for all of the individual 228 populations, but together provide a total of 1,579 individual census estimates from which we can measure 1,353 instances of population growth. These data provide a unique opportunity to study the statistics of human population dynamics across a large sample of traditional human populations bounded geographically over an extended observation window. Moreover, this dataset offers a rare---perhaps unique---opportunity to quantify volatility in traditional societies, and to link observed demographic fluctuations to spatial and social organization. 

Our analysis yields two central results. First, volatility declines with population size as $\sigma(r)\propto N^{-1/4}$, decaying only half as fast as the demographic baseline. Second, we show that in these populations area use scales sublinearly with population size as $A\propto N^{3/4}$, implying that density increases with size and that overlap in individual space use drives correlations in demographic outcomes. Together, these results reveal a density law of demographic stability, $\sigma(r)\propto D^{-1}$, in which volatility declines in direct proportion to density. On average, larger populations are denser and thus more stable than smaller populations, but demographic synchrony means that these populations are always more volatile than would be expected under a simple baseline model of demographic stochasticity.



\section{The model}

\subsection{Definitions and setup}

To illustrate this model we begin by defining variables where $i$ indexes an individual population (society) and $t$ indexes time steps (e.g., years). Let $N_{i,t}$ be a census size of the $i$th population at time $t$. $A_{i,t}$ is then total geographic area associated with that population. It follows then that population density of the $i$th population at time $t$ is $D_{i,t} \equiv N_{i,t}/A_{i,t}$.
The instantaneous growth rate of the $i$th population at time $t$ is then $r_{i,t} \equiv \ln(N_{i,t+1}/N_{i,t})$.

\subsection{Baseline demographic stochasticity}

Under the demographic-noise baseline, individuals are assumed to act independently and the annual probability of an individual reproducing is statistically independent. As such, net demographic increments are approximately Poisson with variance proportional to $N$. Let $\Delta N_{i,t}$ denote the net increment with variance $\mathrm{Var}(\Delta N_{i,t}) = v\,N_{i,t}$. Since $r_{i,t} \approx \Delta N_{i,t}/N_{i,t}$ for small increments, we obtain
\begin{equation}
\mathrm{Var}(r\mid N) \approx \frac{v}{N} \propto N^{-1}, 
\quad \Rightarrow \quad
\sigma(r\mid N) \approx \sqrt{\frac{v}{N}} \propto N^{-1/2}.
\label{eq:baseline}
\end{equation}
Equation~\ref{eq:baseline} is the demographic baseline expectation where demographic volatility decays as $N^{-1/2}$ under demographic independence.

\subsection{A synchrony-extended model}

The observed per-capita growth rate of a population is the average of many individual contributions. Let $y_{ij,t}$ denote the demographic contribution of individual $j$ in population $i$ at time $t$ to the change in population size between $t$ and $t+1$. Each $y_{ij,t}$ is a random variable representing the stochastic outcome of demographic processes such as survival, birth, death, or migration. For example, survival without associated change contributes $y_{ij,t}\approx 0$, death or out-migration contributes $y_{ij,t}<0$, and birth or in-migration associated with individual $j$ contributes $y_{ij,t}>0$.

Formally, we assume
\[
\mathbb{E}[y_{ij,t}] = \mu_i, 
\qquad 
\mathrm{Var}(y_{ij,t}) = s^2,
\]
We introduce pairwise correlation $\rho_{i,t}$ across individuals within the same population:

\[
\mathrm{Corr}(y_{j,t},\,y_{k,t}) \;=\; \rho \quad \text{for } j\neq k.
\]
The per-capita growth rate is then the average of the individual demographic contributions across the population
\begin{equation}
r_{i,t} = \frac{1}{N_{i,t}} \sum_{j=1}^{N_{i,t}} y_{ij,t}.
\label{eq:ydef}
\end{equation}
and the conditional variance of the per-capita growth rate in this synchrony-extended model is then
\begin{equation}
\mathrm{Var}(r\mid N) 
= \frac{1}{N_{i,t}^2}\,\mathrm{Var}\!\Bigg(\sum_{j=1}^{N_{i,t}} y_{ij,t}\Bigg)
= \frac{s^2}{N_{i,t}}\,\big[1+(N_{i,t}-1)\rho_{i,t}\big].
\label{eq:rho}
\end{equation}
If $\rho_{i,t}=0$, we recover the demographic baseline $\sigma(r)\sim N^{-1/2}$ whereas if $\rho_{i,t}>0$, volatility falls more slowly with $N$.

\subsection{Statistical estimation of scaling parameters}
We then wish to use data to estimate the parameters of the volatility scaling law we just derived:
\begin{equation}
\sigma(r \mid N) \;=\; c\,N^{-\alpha}.
\label{eq:pred}
\end{equation}
We proceed as follows. First, population sizes are grouped into exponentially increasing bins along the $N$-axis, with bin sizes $N_{\mathrm{bin}} = 1,2,4,8,16,32,\dots$. Estimating the scaling relation given by equation \ref{eq:pred} requires systematically measuring the change in standard deviation of the growth rate as population increases in scale. Exponential binning of $N$ ensures that each scale of population size is equally represented in the analysis. Because the distribution of $N$ is typically right-skewed, with many small populations and relatively few large ones, equal-width bins would overweight small populations and underweight large ones. Exponential bins balance the representation of different scales, stabilize variance estimates within bins, and ensure that the regression on $\ln N$ is not dominated by the smallest populations. This procedure is standard in scaling analyses in ecology, demography, and statistical physics.

For each bin, we collect all observed growth rates $r_{i,t}$ from populations whose size $N_{i,t}$ falls within that bin, and compute the empirical standard deviation
\[
\hat{\sigma}(r\mid N_{\mathrm{bin}}) \;=\; \sqrt{\mathrm{Var}\{r_{i,t} : N_{i,t}\in N_{\mathrm{bin}}\}}.
\]
This yields a binned set of estimates across the population-size distribution. 
We then fit the log-linearized scaling relation

\begin{equation}
    \ln \hat{\sigma}(r \mid N) \;=\; \ln c \;-\;\alpha \ln N_{\mathrm{bin}} \;+\; \varepsilon,
\end{equation}
where $\ln c$ is the intercept, $-\alpha$ is the slope, and $\varepsilon$ is an error term. Ordinary least squares regression of $\ln \hat{\sigma}(r \mid N)$ on $\ln N$ provides point estimates of $c$ and $\alpha$, while the dispersion of residuals provides confidence intervals. In this framework, $c$ represents the baseline volatility at $N=1$, and $\alpha$ quantifies the rate at which volatility decays with population size.


\section{Results}


\subsection{Data}
We compiled data from the \textit{Insitituto Socioambiental} \cite{isa_pib_downloads} including sources \cite{ricardo1986povos, ricardo1991povos, ricardo1996povos, ricardo2000povos, ricardo2006povos, ricardo2011povos, ricardo2016povos, ricardo2023povos}. These are population census estimates of recently contacted indigenous populations throughout Brazil collected over many generations. These sources provided 1,580 individual census estimates for 228 populations, with the earliest estimate from 1749 and the most recent from 2023.
From these census estimates we were able to calculate 1,353 periods of population growth. For each period of growth we estimated the average annual growth rate over these time windows as

\begin{equation}
\hat {r_{i,t}}=\Bigg(\frac{1}{\Delta t}\Bigg)\ln \!\Bigg(\frac{N_{i,t+\Delta t}}{N_{i,t}} \Bigg)\!.
\label{eq: growth_est}
\end{equation}

\begin{figure}[tbhp]
\centering
\includegraphics[width=1\linewidth]{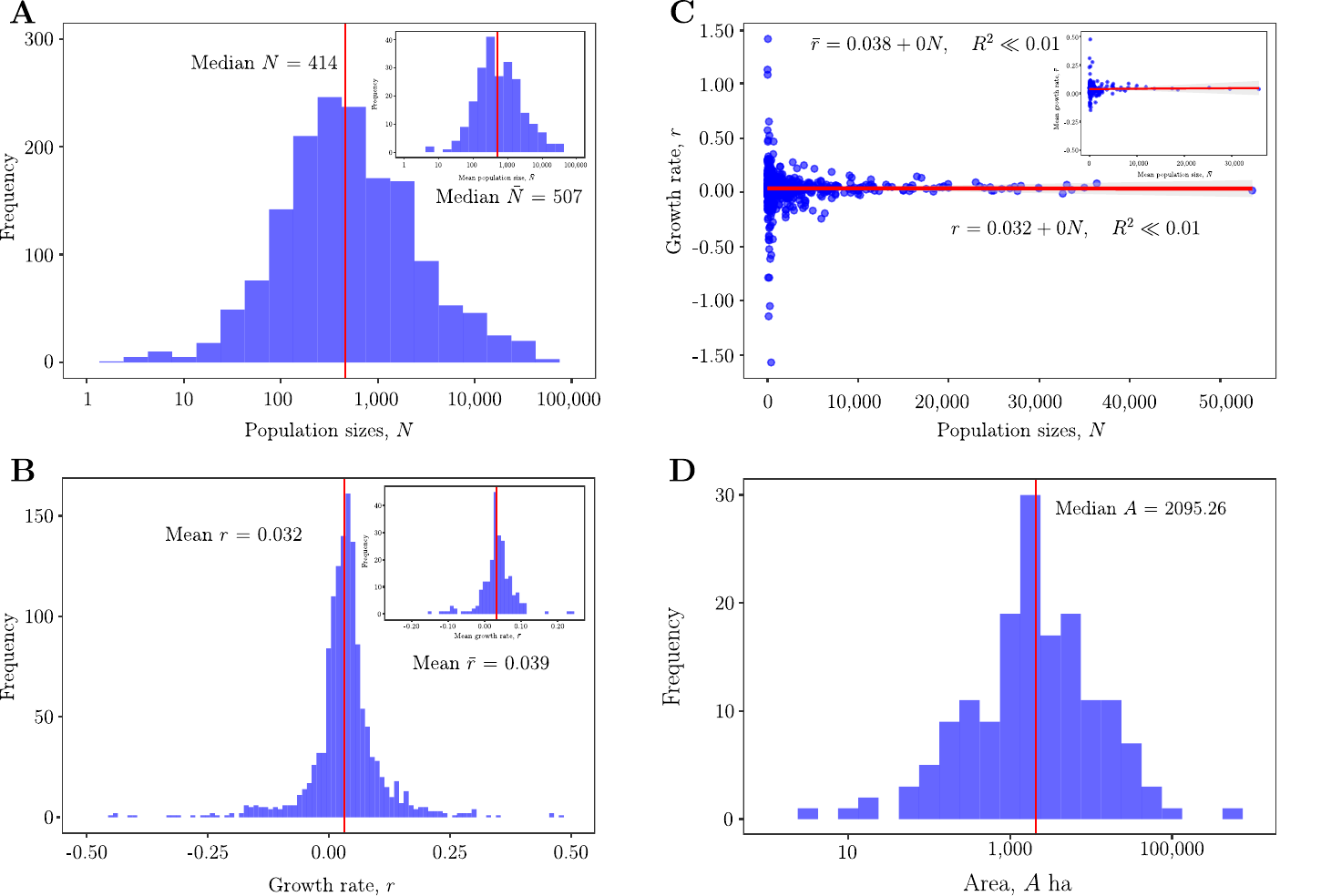}
\caption{Population size and growth rates. A) The distribution of all population size estimates from census populations are approximately lognormal with a median of 414 individuals. Inset in the upper right is the distribution of the average size of the 228 populations over their individual observation windows, with a median size of 507 individuals. B) The distribution of all annual growth rates estimated from the census data (equation \ref{eq: growth_est}) with mean growth rate $\bar r=0.032$ or $3\%$ per year (see Table 1 for details). Inset is the average growth rate for each of the 228 populations over their individual observation windows, with a mean growth rate $\bar r_{N}=0.039$ or $\sim 4\%$ per year. C) Growth rate $r$ as a function of population size $N$ over all census estimates. The relationship shows a funnel plot typical of stochastic population dynamics where population volatility decrease with scale. Inset is the average growth rate of a population $r_N$ as a function of the average population size $\bar N$. In both cases, population growth rates are not size-dependent. D) The distribution of the geographic ranges, or areas $A$ in hectares, for 161 of the 228 populations. The distribution is approximately lognormal with a median size of $2095.26$ ha.}
\label{fig:2x2_plot}
\end{figure}

Figure \ref{fig:2x2_plot}A shows the distribution of population size estimates across the data set and the inset is the average population size for each of the 228 populations. These distributions are approximately lognormal, as might be expected for populations generated by a multiplicative growth process \cite{hamilton2018stochastic}. The median census size over all estimates is 414 and the median across the 228 populations is 507.

Figure \ref{fig:2x2_plot}B shows the distribution of growth rate estimates (equation \ref{eq: growth_est}) across the data set, and the inset is the average growth rate for each of the 228 populations. The distribution is leptokurtic and non-normal though approximately symmetrical. The average growth rate over all estimates is $\bar r=0.032$ or $3.2\%$ annual growth, and the median growth rate across the 228 populations is $r_N=0.039$, or $3.9 \%$.

Figure \ref{fig:2x2_plot}C shows a funnel plot of the growth rate $\hat {r_{i,t}}$ as a function of the corresponding population size ${N_{i,t}}$ at the beginning of the growth period. The inset figure is a plot of the average per population. Fitted OLS regressions (red lines) summarized in Table 1 show that growth rates are independent of populations sizes, but are much more volatile at small population sizes.

\begin{table}[!htbp] \centering 
  \caption{Growth rate and population size scaling} 
  \label{table:rN} 
\scriptsize 
\begin{tabular}{@{\extracolsep{5pt}}lD{.}{.}{-2} } 
\\[-1.8ex]\hline 
\hline \\[-1.8ex] 
 & \multicolumn{1}{c}{\textit{Dependent variable:}} \\ 
\cline{2-2} 
\\[-1.8ex] & \multicolumn{1}{c}{r} \\ 
\hline \\[-1.8ex] 
 $\hat {r_{i,t}}$ & -0.0000$ $(-0.0000$, $0.0000) \\ 
  Constant & 0.03^{***}$ $(0.03$, $0.04) \\ 
 \hline \\[-1.8ex] 
Observations & \multicolumn{1}{c}{1,353} \\ 
R$^{2}$ & \multicolumn{1}{c}{0.0000} \\ 
Residual Std. Error & \multicolumn{1}{c}{0.12 (df = 1351)} \\ 
F Statistic & \multicolumn{1}{c}{0.004 (df = 1; 1351)} \\ 
\hline 
\hline \\[-1.8ex] 
\textit{Note:}  & \multicolumn{1}{r}{$^{*}$p$<$0.1; $^{**}$p$<$0.05; $^{***}$p$<$0.01} \\ 
\end{tabular} 
\end{table} 

Figure \ref{fig:2x2_plot}D shows the distribution of territory size estimates $A$ in hectares (equation \ref{eq: growth_est}) across the data set, and the inset is the average growth rate for each of the 228 populations. The distribution is approximately lognormal. The median territory size across populations is 2095 hectares.

\begin{figure}
\centering
\includegraphics[width=0.8\linewidth]{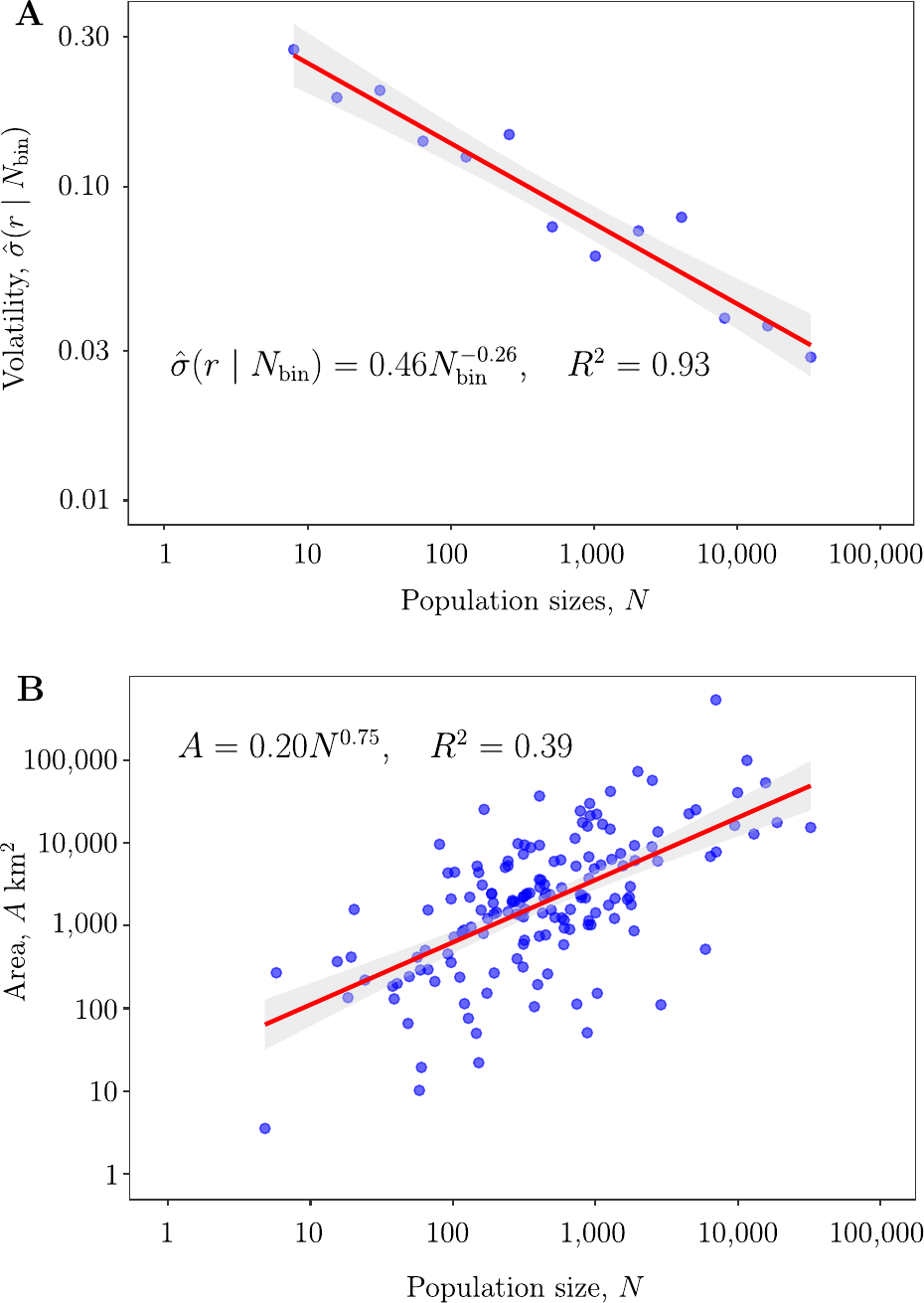}
\caption{Demographic and population size area scaling. A) Observed demographic volatility as a function of binned population size on a log-linear plot, following the methods outlined in the Methods section. Here, volatility declines with population size with a slope $\alpha=-0.26$ (see Table 2 for details). B) The scaling of area use $A$ by population size $N$ for 161 of our 228 populations (limited by data availability) on log-log axes. An OLS estimate of the relationship yields a sublinear slope of $\beta=0.75$ (see Table 3 for details), as is commonly observed in the spatial ecology of traditional societies \cite{hamilton2007nonlinear, hamilton2012human}. }
\label{fig:2_NA_plot}
\end{figure}

Figure \ref{fig:2_NA_plot}A shows the log-log plot of the conditional standard deviation of growth rates--volatilities--$\hat{\sigma}(r\mid N_{\mathrm{bin}})$ as a function of binned population sizes $N_{bin}$.  A fitted OLS regression (red line) summarized in Table 2 shows that across populations, demographic volatility decreases with population size at a rate $\alpha=-0.026$.


\begin{table}[!htbp] \centering 
  \caption{Sigma scaling} 
  \label{table:sigma} 
\scriptsize 
\begin{tabular}{@{\extracolsep{5pt}}lD{.}{.}{-2} } 
\\[-1.8ex]\hline 
\hline \\[-1.8ex] 
 & \multicolumn{1}{c}{\textit{Dependent variable:}} \\ 
\cline{2-2} 
\\[-1.8ex] & \multicolumn{1}{c}{$\hat{\sigma}(r\mid N_{\mathrm{bin}})$} \\ 
\hline \\[-1.8ex] 
 $\alpha$ & -0.26^{***}$ $(-0.30$, $-0.21) \\ 
  $\ln c$ & -0.81^{***}$ $(-1.09$, $-0.53) \\ 
 \hline \\[-1.8ex] 
Observations & \multicolumn{1}{c}{13} \\ 
R$^{2}$ & \multicolumn{1}{c}{0.93} \\ 
Residual Std. Error & \multicolumn{1}{c}{0.20 (df = 11)} \\ 
F Statistic & \multicolumn{1}{c}{142.78$^{***}$ (df = 1; 11)} \\ 
\hline 
\hline \\[-1.8ex] 
\textit{Note:}  & \multicolumn{1}{r}{$^{*}$p$<$0.1; $^{**}$p$<$0.05; $^{***}$p$<$0.01} \\ 
\end{tabular} 
\end{table} 

Figure \ref{fig:2_NA_plot}B shows the log-log plot of the area of the territory size of a population as a function of the population size.  A fitted OLS regression (red line) summarized in Table 3 shows that across populations, area scales positively with population size at a rate $\beta=0.75$.

\begin{table}[!htbp] \centering 
  \caption{Area-population size scaling} 
  \label{} 
\scriptsize 
\begin{tabular}{@{\extracolsep{5pt}}lD{.}{.}{-2} } 
\\[-1.8ex]\hline 
\hline \\[-1.8ex] 
 & \multicolumn{1}{c}{\textit{Dependent variable:}} \\ 
\cline{2-2} 
\\[-1.8ex] & \multicolumn{1}{c}{lnA} \\ 
\hline \\[-1.8ex] 
 $\beta$ & 0.75^{***}$ $(0.61$, $0.90) \\ 
  Constant & 2.99^{***}$ $(2.07$, $3.90) \\ 
 \hline \\[-1.8ex] 
Observations & \multicolumn{1}{c}{160} \\ 
R$^{2}$ & \multicolumn{1}{c}{0.39} \\ 
Residual Std. Error & \multicolumn{1}{c}{1.45 (df = 158)} \\ 
F Statistic & \multicolumn{1}{c}{100.94$^{***}$ (df = 1; 158)} \\ 
\hline 
\hline \\[-1.8ex] 
\textit{Note:}  & \multicolumn{1}{r}{$^{*}$p$<$0.1; $^{**}$p$<$0.05; $^{***}$p$<$0.01} \\ 
\end{tabular} 
\end{table} 



\section{Model analysis}
\subsection{Synchrony scaling with $N$}

Table 1 shows $\alpha=-0.26 \pm 0.05 \approx -1/4$ and so empirically $\sigma(r) \propto N^{-1/4} \Rightarrow \mathrm{Var}(r)\propto N^{-1/2}$. As such, volatility decays at half the rate expected by the baseline demographic model. For equation (\ref{eq:rho}) to produce this result, synchrony (pairwise correlations) must decline with size as
\begin{equation}
\rho(N) \ \propto\ N^{-1/2}.
\label{eq:rhoN}
\end{equation}
Then for large $N$, the synchrony term dominates:
\begin{equation}
\mathrm{Var}(r\mid N) \approx s^2\,c_\rho N^{-1/2}, 
\quad \Rightarrow \quad \sigma(r\mid N)\propto N^{-1/4},
\label{eq:empirical}
\end{equation}
matching the results we obtain from the data.

Further, these results suggest populations consist of effective correlated ``clusters'', or modular demographic units of average size $m(N)\propto N^{1/2}$. The effective number of independent draws is $K(N)\propto N^{1/2}$, yielding $\sigma(r)\sim K^{-1/2}\sim N^{-1/4}$. As such, volatility scales negatively with population size at half the rate of the baseline demographic model as pairwise correlations within populations dampen the rate at which stochasticity is averaged out by scale.

\subsection{Link to space use and density}

Table 2 shows $\beta=3/4$. Since $A\propto N^{3/4}$, we have $A/N\propto N^{-1/4}$ and so $D=N/A\propto N^{1/4}$. Substituting into equation (\ref{eq:empirical}) we have,
\begin{equation}
\sigma(r) \ \propto\ N^{-1/4}\ =\ (N/A)^{-1}\ =\ D^{-1}.
\end{equation}
Thus, stability increases in direct proportion to density as doubling density halves volatility.




\subsection{Unified variance model}

Conceptually, we can then define a full growth model composed of three terms:

\begin{equation}
   r = f(\mbox{individual contributions, demography, synchrony}) 
\end{equation}

We then formalize this into a general model that considers the effects of demography and synchrony (pairwise correlations) on population growth:
\begin{equation}
r_{i,t} = \mu_i - \phi \log N_{i,t} 
+ \sigma_d N_{i,t}^{-1/2}\eta_{i,t}
+ \sigma_\rho N_{i,t}^{-1/4}\xi_{i,t},
\label{eq:fullmodel}
\end{equation}


where $\phi$ is a density-dependence coefficient and $\eta$ and $\xi\sim\mathcal{N}(0,1)$. The conditional variance is then
\begin{equation}
\mathrm{Var}(r\mid N) = \sigma_d^2{N^{-1}} + \sigma_\rho^2 N^{-1/2}.
\label{eq:fullvar}
\end{equation}


\subsection{Implications: synchrony inflates volatility}

\begin{figure}[tbhp]
\centering
\includegraphics[width=0.5\linewidth]{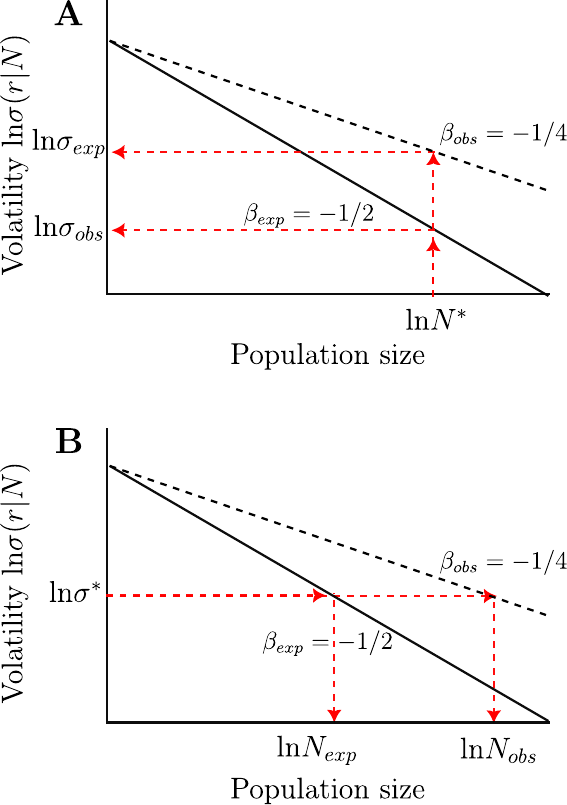}
\caption{Demographic fluctuations schematic. A) For a given population size, $N^*$, the expected volatility of the population $\sigma_{exp}$ is lower than the observed volatility $\sigma_{obs}$ as the observed slope $\beta_{obs}=-1/4$ is shallower than the expected slope $\beta_{obs}=-1/2$. B) For a given level of volatility, $\sigma^*$, the expected population size $N_{exp}$ is lower than the observed population size $N_{obs}$. As such, all populations are more volatile than expected for any given size, or equivalently, large populations are observed to be as volatile as smaller populations under the expected baseline demographic model.}
\label{fig:3_schematic_plot}
\end{figure}

Equations (\ref{eq:rho})--(\ref{eq:fullmodel}) show that for a given population size increasing $\rho$ inflates $\mathrm{Var}(r)$, because correlated outcomes accumulate rather than cancel. Populations of size $N$ behave as if they had only $K\propto N^{1/2}$ independent individuals, increasing demographic risk, as $K(N)=N/m(N)=N \times N^{-1/2}=\sqrt{N}$. As such, in synchronous populations the effective demographic degrees of freedom are vastly reduced (Figure \ref{fig:3_schematic_plot}A).
But across populations empirically, $\rho(N)\propto N^{-1/2}$, so synchrony declines with density and size, producing the observed $\sigma(r)\propto D^{-1}$.
Thus, demographic synchrony increases volatility, but synchrony itself declines across populations as population density increases.



The observed scaling of volatility and space use shows that human demography is not reducible to independent births and deaths but is structured by social, ecological, and institutional synchrony. Synchrony inflates volatility, but density, sharing, and spatial organization counteract synchrony, stabilizing demographic trajectories.
Table~\ref{tab:exponent-mapping} summarizes this chain of exponents linking space use, correlation, and volatility. 

\begin{table}[t]
\centering
\small
\caption{Exponent mapping from spatial overlap to correlation and volatility.}
\begin{tabularx}{\textwidth}{lXc}
\hline
\textbf{Quantity} & \textbf{Scaling relation} & \textbf{Implied exponent} \\
\hline
Area use (per capita) & $\displaystyle A/N \sim N^{-1/4}$ & $-1/4$ \\
Population density & $\displaystyle D = N/A \sim N^{1/4}$ & $1/4$ \\
Pairwise correlation (synchrony) & $\displaystyle \rho(N) \sim N^{-1/2}$ & $-1/2$ \\
Variance of growth rate & $\displaystyle \mathrm{Var}(r \mid N) \sim N^{-1/2}$ & $-1/2$ \\
Volatility (std. dev.) & $\displaystyle \sigma(r) = N^{-1/4}$ & $-1/4$ \\
Density law of volatility & $\displaystyle \sigma(r) \sim D^{-1}$ & $-1$ \\
\hline
\end{tabularx}
\label{tab:exponent-mapping}
\end{table}

\section{Discussion}

Using census data, we show that demographic synchrony at the individual scale increases volatility at the population scale meaning that even large-scale human populations are prone to stochastic demographic fluctuations. As such, population size alone does not ensure demographic stability as much as it would do if the demographic outcomes of individuals were statistically independent, as usually assumed. Classical demography predicts that volatility in human population growth scales as $\sigma(r)\propto N^{-1/2}$, but empirically we find $\sigma(r)\propto N^{-1/4}$. This departure from the baseline of independent demographic stochasticity implies that individuals within populations do not behave as statistically independent demographic units, dampening the otherwise stabilizing effect of the law of large numbers. Instead, the demographic outcomes of individuals are correlated reducing the effective degrees of freedom in the population. Consequently, a community of $N=400$ individuals behaves demographically as though it contained only $K(N)\approx 20$ demographically independent units. This dramatic reduction explains why volatility remains systematically higher than expected and why large-scale populations remain vulnerable to stochastic events.

In our model, pairwise correlation $\rho$ is the mechanism through which independence is broken and volatility is inflated. In anthropological terms, however, these correlations are not just abstract statistical parameters but the demographic imprint of social and ecological embedding. For example, kinship, reciprocity, and inter-household coordination cluster births, deaths, and reproductive outcomes through institutions and norms that coordinate childcare, food production, resource sharing, and subsistence, all of which are well-studied \cite{hill2009cooperative, hill2017ache, gurven2000s, page2019testing, kramer2010cooperative, gurven2004give, kraft2021energetics, kraft2023female, hill2002altruistic}. 
Mobility and fission--fusion dynamics synchronize outcomes further at multiple scales, as mobility rarely occurs in isolation but usually involves modular groupings of families or co-resident camping groups not only moving across landscapes but through social networks \cite{hill2011co, hill2014hunter, hamilton2016ecological, hamilton2024food}. Epidemiological exposure adds another layer as dense social interactions structure transmission networks in which disease outbreaks synchronize mortality across households, as starkly illustrated by catastrophic waves of measles and influenza epidemics throughout post-contact Amazonia \cite{hemming1978red, vaz2011isolados, ribeiro1967indigenous, hurtado2001epidemiology, bodard1974green, hamilton2014crash}.

The synchrony-extended model captures these processes by introducing $\rho$ among individuals. When $\rho=0$, volatility declines as $\sigma(r)\sim N^{-1/2}$; when $\rho>0$, off-diagonal covariance terms slow the decay, producing the empirical scaling $\sigma(r)\sim N^{-1/4}$. This scaling requires $\rho(N)\sim N^{-1/2}$, implying that pair-wise correlations weaken with size but more slowly than expected under independence. The result is modular structure: effective demographic cluster size grows as $m(N)\sim \sqrt{N}$. Synchrony therefore systematically dampens independence, ensuring elevated volatility across scales.

The scaling of area use, $A\sim N^{3/4}$, provides a clear ecological mechanism for this synchrony. Because per-capita area shrinks as $A/N\sim N^{-1/4}$, individual space use is increasingly shared--and thus coordinated--as populations grow meaning that larger populations are also denser; $D=N/A\sim N^{1/4}$. Increasing spatial overlap results in increasing shared exposure to environmental fluctuations synchronizing demographic outcomes across households. At the same time, higher population density creates increased opportunities for buffering through social mechanisms such as kinship, exchange, and sharing that redistribute environmental shocks across households, reducing one-to-one correlations. This duality yields the density law $\sigma(r)\propto D^{-1}$, with volatility declining as density rises. 

Thus, population size does decrease the volatility of populations but it does so much more slowly than would be expected if there were no demographic synchrony. In other words, larger populations are more stable than smaller populations, but not as much as would be expected under a baseline model. Large populations are more stable as they are on average denser and density breaks synchrony; $\rho\sim D^{-2}$. So, scale still stabilizes synchronous populations via increasing population density, but synchrony ensures that stability grows only slowly with size, leaving large populations more volatile than classical demographic theory predicts.

The implications of these findings are twofold. First, sparse populations are especially vulnerable as low population density heightens synchrony and leaves fewer buffering mechanisms. Second, dense populations, though more stable on average, remain vulnerable. Density decreases volatility but it does not remove it. 
This duality clarifies the fundamental relationship between resilience and fragility in human populations. Cooperation evolved to buffer individuals against everyday risk by pooling labor, food, and resources, stabilizing individual outcomes, but, by definition, cooperation synchronizes outcomes, inflating population-level volatility. We term this the cooperation–synchrony paradox: the very strategies that evolved to buffer individuals against stochastic sources of risk have the effect of reducing demographic independence at the population level, making societal collapse more likely. This paradox explains why human populations remain demographically fragile at all scales, even as cooperative institutions expand \cite{hamilton2024institutional}.

The consequences for extinction risk are substantial. Classical population viability analyses assume $\sigma^2(r)\propto 1/N$, predicting relatively low minimum viable population sizes (MVPs) \cite{morris2002quantitative, dennis1991estimation, traill2007minimum}. Under synchrony, however, minimum viable population sizes are roughly the square of baseline estimates. A population predicted to persist at $N_{\text{MVP}}=100$ under independence may require $10{,}000$ under synchrony. Populations that appear large enough to be stable may still be vulnerable (Figure \ref{fig:3_schematic_plot}B).


Although our empirical analyses draw on a uniquely rich dataset of indigenous Amazonian populations, the results are not limited to this regional context. Rather, we argue they reveal general demographic principles that apply broadly to small-scale traditional societies. Wherever they have been measured and quantified, ethnographic societies display remarkably similar structural properties of social and ecological embedding, overlapping land use, modular social structures and fission--fusion dynamics. These are the features that generate correlated demographic fates, or synchrony, we observe among Amazonian groups.

The synchrony-extended model formalizes a mechanism that is not culture- or region-specific. When demographic events are correlated across individuals or households, the stabilizing effect of size is weakened, volatility declines more slowly with $N$, and extinction risk increases. This mathematical principle should thus hold for all foraging, horticultural, pastoral, and small-scale agrarian systems alike, and perhaps human population dynamics in general. In all such contexts, dense social networks and shared ecologies couple demographic trajectories, reducing effective independence and inflating variance relative to the baseline expectation of demographic stochasticity. Clearly, this is a hypothesis to be tested wherever data are available. The Amazonian case is therefore not an anomaly, but an extremely rare empirical window into a general demographic rule. These data provide a unique quantitative test of how cooperation, spatial overlap, and ecological coupling interact to structure demographic variability. We expect similar scaling relationships to characterize other small-scale populations where local cooperation and ecological interdependence synchronize demographic outcomes.

\section{Conclusion} 
While demographic synchrony increases population volatility, it does not eliminate the stabilizing role of scale altogether; it simply dampens its effects. Even when individual demographic outcomes are correlated, larger populations remain more stable than smaller ones, because larger populations are also more dense as $D\propto N^{1/4}$. 
Recasting the volatility law as $\sigma(r)\propto D^{-1}$ makes this clear: increasing density, which accompanies increasing $N$, systematically reduces volatility but this stabilizing effect is weaker than the classical expectation under demographic independence. 
Thus, while scale increases stability, it does so much more slowly than demographic theory predicts under independence with the consequence that even large populations remain volatile and vulnerable to collapse despite their size. 

While demographic synchrony amplifies population volatility, it does not erase the stabilizing effects of scale, it simply dampens them. Even when individual demographic outcomes are correlated, larger populations remain more stable than smaller ones because population density increases with size as $D\propto N^{1/4}$. Recasting the volatility law as $\sigma(r)\propto D^{-1}$ makes this relationship explicit: increasing density systematically reduces volatility, but the stabilizing effect of scale is weaker than predicted under demographic independence. In other words, synchrony imposes a limit on how much stability scale alone can confer.

Thus, while increasing scale still enhances demographic resilience, it does so far more slowly than classical theory would expect. The practical consequence is that even relatively large populations that, by size alone, should be buffered against random fluctuations remain volatile and vulnerable to collapse when internal dynamics become synchronized. This finding highlights an understudied general principle of human demography: correlation, not merely size, governs stability. Systems that grow without mechanisms to desynchronize their internal fluctuations risk instability regardless of scale.

From small-scale foragers to large-scale empires, demographic synchrony is inescapable. In small-scale societies, synchrony arises from overlapping land use, kinship, and food sharing. In large-scale societies, synchrony emerges from common reliance on staple crops, market integration, centralized institutions, and exposure to regional shocks. Extending the ecological concept of synchrony to human demography yields a general principle: correlations reduce effective degrees of freedom, inflating variance and raising extinction risk. 



\pagebreak

\bibliographystyle{unsrt}
\bibliography{amazon}


\end{document}